\documentclass{aa}
\usepackage{graphics}
\usepackage{psfig}
\def\h2{H{\small II}}
\begin{document}
%-----------------------------------------------------------------
\title{Discovery of the high--ionization emission line [Ne V] $\lambda$3426 
in the blue compact dwarf galaxy Tol 1214--277\thanks{Based on observations
collected at the European Southern Observatory, Chile, ESO program 
71.B-0032(A).}}
%
%\subtitle{A Clue to Dwarf Galaxy Evolution?}
%
\author{Y. I.\ Izotov \inst{1}
\and K. G.\ Noeske\inst{2,}\inst{3}
\and N. G.\ Guseva \inst{1}
\and P.\ Papaderos \inst{2}
\and T. X.\ Thuan \inst{4}
\and K. J.\ Fricke   \inst{2}}
\offprints{Y.I. Izotov, izotov@mao.kiev.ua}
\institute{      Main Astronomical Observatory,
                 Ukrainian National Academy of Sciences,
                 Zabolotnoho 27, Kyiv 03680,  Ukraine
\and
                 Universit\"ats--Sternwarte, Geismarlandstra\ss e 11,
                 D--37083 G\"ottingen, Germany
\and
                 University of California, 1156 High St., Santa Cruz,
                 CA 95064, USA
\and
                 Department of Astronomy, University of Virginia, 
                 Charlottesville, VA 22903, USA
%\and
%                 National Optical Astronomy Observatory, 
%                 Tucson, AZ 85726, USA
}

\date{Received \hskip 2cm; Accepted}
%\date{Received(00.00.0000)/Accepted(00.00.0000)}
%\titlerunning{High--ionization emission in Tol 1214--277}
%\authorrunning{Y. I. Izotov et al.}
%\maketitle

% ======================================================================================
\abstract{The discovery of the high-ionization [Ne {\sc v}] $\lambda$3426\AA\ 
emission line in the spectrum of the blue compact dwarf (BCD) 
galaxy Tol 1214--277 is reported. The detection of this
line implies the presence of intense ionizing X-ray emission 
with a luminosity $L_{\rm x}$ in the range 10$^{39}$ -- 10$^{40}$ erg s$^{-1}$.
Such a high X-ray luminosity cannot be reproduced by models 
of massive stellar populations.
Other mechanisms, such as fast shocks or accretion of gas in high-mass X-ray 
binaries need to be invoked to account for the high intensity of the 
[Ne {\sc v}] $\lambda$3426\AA\ emission line.
}

\maketitle
%---------------------------------------------------------------------------------------------

\markboth {Y.I. Izotov et al.}{High--ionization emission 
line [Ne {\sc v}]$\lambda$3426 in Tol 1214--277}
%----------------------------------------------------------------------------------------------
\section {Introduction}
\label{intro}

Tol 1214--277 is one of the best studied blue compact dwarf (BCD) galaxies
in the Southern sky. Numerous spectroscopic observations of its brightest
H {\sc ii} region
(e.g., Kunth \& Sargent \cite{Kunth83}; Campbell, Terlevich \& Melnick
\cite{Cam86}; Pagel et al. \cite{Pag92}; Masegosa, Moles \& Campos-Aguilar
\cite{Mas94}; Fricke et al. \cite{Fricke01}; Izotov, Chaffee \& Green
\cite{ICG01}) have revealed the
very low oxygen abundance changing in the range of 12 + log O/H = 7.51 -- 7.55.
%or $\sim$ $Z_\odot$/25 adopting the solar 12 + log O/H = 8.92 from
%Anders \& Grevesse (\cite{Anders89}). 
This makes Tol~1214--277 a possible candidate for being a young unevolved 
galaxy (Fricke et al. \cite{Fricke01}), suitable for the primordial helium 
abundance determination 
(Pagel et al. \cite{Pag92}; Izotov, Chaffee \& Green \cite{ICG01}).

The spectrum of the brightest H {\sc ii} region in Tol~1214--277 is 
characterised by strong nebular emission lines suggesting a very young
age ($\sim$ 3 Myr) for the ionizing cluster. A strong He {\sc ii} $\lambda$4686
emission line is a common property of low-metallicity BCDs 
(Guseva, Izotov \& Thuan \cite{GIT00}).
Its presence in the spectrum of Tol 1214--277 implies the existence
 of hard ionizing radiation with $\lambda$ $<$ 228\AA. However, the 
He {\sc ii} $\lambda$4686 emission line in Tol 1214--277 is the strongest 
among known BCDs, reaching $\sim$ 5\% of the H$\beta$ emission line flux and
suggesting that the hard radiation in this galaxy is particularly intense.
Further evidence for the presence of the hard radiation in Tol 1214--277 was
found by Fricke et al. (\cite{Fricke01}) who discovered the 
[Fe~{\sc v}]~$\lambda$4227\AA\ emission line in its spectrum. Until now, 
this line has been definitely detected in only two BCDs,
Tol 1214--277 and SBS 0335--052 (Fricke et al. \cite{Fricke01}; 
Izotov, Chaffee \& Schaerer \cite{ICS01}). [Fe {\sc v}] emission
can be present only in the He$^{+2}$ zone of the H {\sc ii} region. 
The ionization potential of the Fe$^{+3}$ ion is 4.028 Rydberg and hence the
Fe$^{+4}$ ion can be produced only by radiation with $\lambda$$\la$200\AA.
Fricke et al. (\cite{Fricke01}) and Izotov et al.(\cite{ICS01}) have discussed
different mechanisms which can be responsible for 
the hard radiation in Tol 1214--277 and SBS 0335--052. They 
concluded that ionizing stellar radiation is too soft to explain the 
strong He {\sc ii} $\lambda$4686 and [Fe {\sc v}] $\lambda$4227
emission lines. Other ionization sources, such as fast shocks and high-mass  
X-ray binary systems, need to be considered.

In this paper we report the discovery of the [Ne~{\sc v}]~$\lambda$3426 
emission line in the spectrum of Tol 1214--277 based on new 
spectroscopic observations. This finding further supports the presence of 
a very highly ionized gas component in the brightest H {\sc ii} region of 
Tol 1214--277.

%*************************************************************
%             Fig.1
%*************************************************************
\begin{figure}[hbtp]%[tbh]
%\vspace{7.cm}
%\hspace*{1.0cm}
\psfig{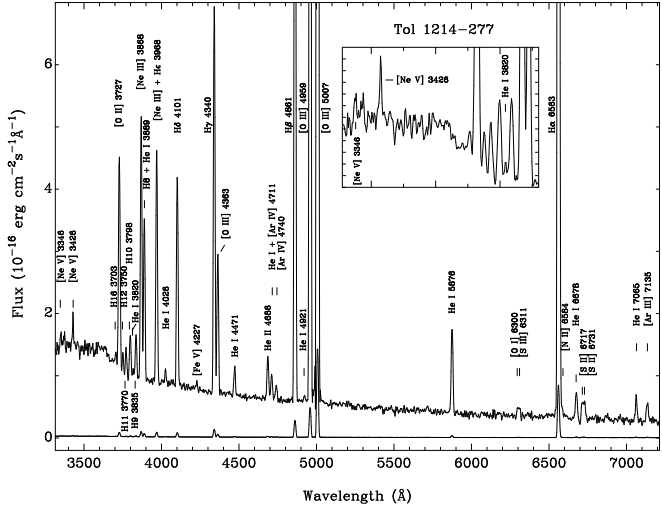}
    \caption{The redshift-corrected spectrum of the brightest H {\sc ii} 
region in Tol 1214--277 with labelled emission lines. The lower spectrum is 
the observed spectrum downscaled by a factor of 50. The inset shows a
close-up view of the blue part of the spectrum with the two high-ionization 
[Ne {\sc v}] $\lambda$3346\AA\ and [Ne~{\sc v}]~$\lambda$3426\AA\ emission
lines.}
    \label{f:bright}
\end{figure}
%----------------------------------------------------------------------------------------

% =======================================
\section{Observations and data reduction \label{obs}} 

% The long-slit spectroscopic observation was carried out on 
%20th April 1994 with the 
%4.5m  Multiple Mirror telescope (MMT)  over a wavelength range 3650 -- 7500\AA.

Long-slit spectroscopic observations of Tol 1214--277 were carried out 
during two nights, on 24 and 25 April, 2003 with the ESO 3.6-m telescope 
(La Silla) in conjunction with the EFOSC spectrograph. 
The long slit with width of 1\arcsec\ was centered on the brightest H {\sc ii} 
region at a position angle P.A. = --34\fdg2. The grism $\#11$ was
used giving the wavelength range $\lambda$3400--$\lambda$7400 and the spectral
resolution of $\sim$13.2~\AA\ (FWHM). The spatial scale along the slit of 
0\farcs157 pixel$^{-1}$ was binned by a factor of 2 resulting in the spatial
resolution of 0\farcs314 pixel$^{-1}$. The total exposure of the 
Tol 1214--277 observations was 80 min during two nights, splitted 
into four subexposures.
No correction for atmospheric refraction was made because of 
the small airmass $\sim$ 1.003 during the observations.

The data reduction was made with IRAF\footnote{IRAF is 
the Image Reduction and Analysis Facility distributed by the 
National Optical Astronomy Observatory (NOAO), which is operated by the 
Association of Universities for Research in Astronomy (AURA) under 
cooperative agreement with the National Science Foundation (NSF).} 
software package. This includes  bias--subtraction, 
flat--field correction, cosmic-ray removal, wavelength calibration, 
night sky background subtraction, correction for atmospheric extinction and 
absolute flux calibration of the two-dimensional spectrum.

An one-dimensional spectrum of the brightest H {\sc ii} region 
was extracted within an aperture of 
1\arcsec\ $\times$ 2\farcs6. It was corrected for the redshift
$z$ = 0.02592 $\pm$ 0.00017 which is derived from the observed wavelengths 
of 28 strongest emission lines. The spectrum is shown in Fig.~\ref{f:bright}
and is characterised by strong nebular emission lines and the Balmer jump at 
$\sim$ $\lambda$3660\AA. 
%The most striking feature of the spectrum is the
%presence of strong [Ne {\sc v}] $\lambda$3426\AA\ emission and possibly 
%[Ne {\sc v}] $\lambda$3346\AA\ emission.
%The inset in Fig.~\ref{f:bright} shows a close-up view
%of the blue part of the spectrum containing the Balmer jump
%and the [Ne {\sc v}] emission lines.

%--------------------------------------------------------------
%     Table 1a
%--------------------------------------------------------------

\begin{table}[tbh]
%     \centering{
\caption{Fluxes and equivalent widths of emission lines in the 
H~{\sc ii} region.}
\label{t:Intens}
\begin{tabular}{lcr} \hline \hline
%  &\multicolumn{3}{c}{MMT}&&\multicolumn{3}{c}{ESO  (3.6m)} \\ \cline{2-4} \cline{6-8}
$\lambda_{0}$(\AA) Ion                  
&$I$($\lambda$)/$I$(H$\beta$)&$EW$(\AA)  \\ \hline
3346\ [Ne {\sc v}]            & 0.013 $\pm$0.007 & 1.4 $\pm$0.7    \\
3426\ [Ne {\sc v}]            & 0.027 $\pm$0.004 & 2.9 $\pm$0.4    \\
3704\ H16                     & 0.033 $\pm$0.012 & 1.6 $\pm$0.4    \\
3727\ [O {\sc ii}]            & 0.280 $\pm$0.006 &40.6 $\pm$0.5    \\
3750\ H12                     & 0.038 $\pm$0.007 & 3.15 $\pm$0.4   \\
3770\ H11                     & 0.052 $\pm$0.006 & 5.9 $\pm$0.5    \\
3798\ H10                     & 0.071 $\pm$0.005 & 9.2 $\pm$0.5    \\
3820\ He {\sc i}              & 0.004 $\pm$0.002 & 0.7 $\pm$0.2    \\
3835\ H9                      & 0.067 $\pm$0.005 & 8.4 $\pm$0.5   \\
3868\ [Ne {\sc iii}]          & 0.320 $\pm$0.006 & 56.1 $\pm$0.5   \\
3889\ H8 + He {\sc i}         & 0.210 $\pm$0.005 &34.7 $\pm$0.6   \\
3968\ [Ne {\sc iii}] + H7     & 0.303 $\pm$0.006 &56.3 $\pm$0.6   \\
4026\ He {\sc i}              & 0.015 $\pm$0.003 & 2.9 $\pm$0.5   \\
%4068\ [S {\sc ii}]           &       ...        &\multicolumn {1}{c}{...}  \\ 
4101\ H$\delta$               & 0.266 $\pm$0.005 &51.4 $\pm$0.4   \\
4227\ [Fe {\sc v}]            & 0.009 $\pm$0.002 & 2.0 $\pm$0.4   \\
4340\ H$\gamma$               & 0.474 $\pm$0.008 &114.7$\pm$0.6   \\
4363\ [O {\sc iii}]           & 0.166 $\pm$0.003 &41.5 $\pm$0.4   \\
4471\ He {\sc i}              & 0.030 $\pm$0.002 & 8.0 $\pm$0.4   \\
4686\ He {\sc ii}             & 0.050 $\pm$0.001 &14.7 $\pm$0.4   \\
4713\ [Ar {\sc iv}] + He {\sc i} & 0.029 $\pm$0.001 & 8.6 $\pm$0.4   \\
4740\ [Ar {\sc iv}]           & 0.017 $\pm$0.001 & 5.1 $\pm$0.4   \\
4861\ H$\beta$                & 1.000 $\pm$0.015 &320.4$\pm$0.8   \\
4922\ He {\sc i}              & 0.006 $\pm$0.001 & 2.2 $\pm$0.4   \\
4959\ [O {\sc iii}]           & 1.734 $\pm$0.026 &572.7$\pm$1.0   \\
5007\ [O {\sc iii}]           & 5.219 $\pm$0.076 &1750.0$\pm$1.8  \\
%5199\ [N {\sc i}]            & 0.005 $\pm$0.001 &1.0 $\pm$0.2  \\
%5271\ [Fe {\sc iii}]         & 0.003 $\pm$0.001 & 0.7 $\pm$0.2 \\
%5518\ [Cl {\sc iii}]         & 0.003 $\pm$0.001 & 0.8 $\pm$0.2 \\
%5538\ [Cl {\sc iii}]         & 0.002 $\pm$0.001 & 0.4 $\pm$0.2 \\
5876\ He {\sc i}              & 0.092 $\pm$0.002 &45.9 $\pm$0.8   \\
6300\ [O {\sc i}]             & 0.006 $\pm$0.001 & 4.3 $\pm$0.5   \\
6312\ [S {\sc iii}]           & 0.006 $\pm$0.001 & 4.2 $\pm$0.4   \\
%6364\ [O {\sc i}]            & 0.008 $\pm$0.001 & 2.7 $\pm$0.2 \\
6563\ H$\alpha$               & 2.752 $\pm$0.044 &1737.0$\pm$2.6  \\
%6584\ [N {\sc ii}]           & 0.016 $\pm$0.002 & 9.9 $\pm$1.3   \\
6678\ He {\sc i}              & 0.027 $\pm$0.001 &19.3 $\pm$0.9   \\
6717\ [S {\sc ii}]            & 0.015 $\pm$0.001 &10.1 $\pm$0.8   \\
6731\ [S {\sc ii}]            & 0.013 $\pm$0.001 & 9.5 $\pm$1.0   \\
%6717+31\ [S {\sc ii}]         & 0.035 $\pm$0.001 &26.1 $\pm$1.2   \\  
7065\ He {\sc i}              & 0.023 $\pm$0.001 &19.6 $\pm$1.0   \\
7135\ [Ar {\sc iii}]          & 0.021 $\pm$0.001 &18.8 $\pm$1.2   \\
%                     &  &  \\
$C$(H$\beta$)\ dex             &\multicolumn {2}{c}{0.220$\pm$0.019}  \\
$F$(H$\beta$)$^{\rm a}$        &\multicolumn {2}{c}{1.89$\pm$0.01}    \\
EW(abs)~\AA                  &\multicolumn {2}{c}{2.9$\pm$0.4}      \\
\hline\hline
\end{tabular}

$^{\rm a}$in units 10$^{-14}$\ erg\ s$^{-1}$cm$^{-2}$.
%}
\end{table}
%-------------------------------------------------------------

%--------------------------------------------------------------------------------
%    Table 2
%-------------------------------------------------

\begin{table}[tbh]
\begin{center}
\caption{Derived physical parameters of the H {\sc ii} region.}
\label{t:Chem}
\begin{tabular}{lcc} \hline \hline
Parameter                      &  Value \\ \hline
$T_{\rm e}$(O {\sc iii})(K)               &19380$\pm$250  \\
$T_{\rm e}$(O {\sc ii})(K)                &15500$\pm$180  \\
$T_{\rm e}$(Ar {\sc iii})(K)              &17780$\pm$210  \\
$T_{\rm e}$(S {\sc iii})(K)               &17780$\pm$210  \\ 
$N_{\rm e}$(S {\sc ii})(cm$^{-3}$)        & 380$\pm$260    \\ %\\
%$N_{\rm e}$(He {\sc ii})(cm$^{-3}$)             &    ..$\pm$..  
%$\tau$($\lambda$3889)               &       0.0     
O$^+$/H$^+$($\times$10$^5$)         &0.229$\pm$0.008 \\
O$^{+2}$/H$^+$($\times$10$^5$)      &3.190$\pm$0.100 \\
O$^{+3}$/H$^+$($\times$10$^5$)      &0.189$\pm$0.011 \\
O/H($\times$10$^5$)                 &3.609$\pm$0.101 \\
12 + log(O/H)                       &7.56$\pm$0.01    \\ %\\
%---------------------------------------------------------------------------
%N$^{+}$/H$^+$($\times$10$^7$)       &1.305$\pm$0.375 \\
%ICF(N)$^{\rm a}$                          &20.62~~~~~~~~~~~\\
%log(N/O)                             &--1.326$\pm$0.179~~\\ \\
%---------------------------------------------------------------------------
Ne$^{+2}$/H$^+$($\times$10$^5$)     &0.401$\pm$0.013 \\
ICF(Ne)$^{\rm a}$                         &1.13\,~~~~~~~~~~~\\
log(Ne/O)                            &--0.90$\pm$0.02~~\\ %\\
%---------------------------------------------------------------------------
S$^+$/H$^+$($\times$10$^7$)         &0.273$\pm$0.018 \\
S$^{+2}$/H$^+$($\times$10$^7$)      &1.915$\pm$0.211 \\
ICF(S)$^{\rm a}$                          &3.45\,~~~~~~~~~~~\\
log(S/O)                             &--1.68$\pm$0.04~~\\ %\\
%---------------------------------------------------------------------------
%Cl$^{++}$/H$^+$($\times$10$^8$)      &1.017$\pm$0.223  \\
%ICF(Cl)$^{\rm a}$                          &1.57~~~~~~~~~~~~~~\\
%log(Cl/O)                             &--3.409$\pm$0.076~~\\ \\
%-----------------------------------------------------------------------
Ar$^{+2}$/H$^+$($\times$10$^7$)        &0.590$\pm$0.039 \\
Ar$^{+3}$/H$^+$($\times$10$^7$)      &1.404$\pm$0.105 \\
ICF(Ar)$^{\rm a}$                          &1.00\,~~~~~~~~~~~\\
log(Ar/O)                             &--2.26$\pm$0.03~~\\ %\\ %\hline
%----------------------------------------------------------------------------
% for He abundance it is used the data from Smith, 1996 for 3.6m data
%He$^+$/H$^+$($\lambda$4471)         &0.066$\pm$0.003 \\
%He$^+$/H$^+$($\lambda$5876)         &0.078$\pm$0.002 \\
%He$^+$/H$^+$($\lambda$6678)         &0.083$\pm$0.004: \\
He$^+$/H$^+$  (weighted mean)       &0.076$\pm$0.001 \\
He$^{+2}$/H$^+$ (from He {\sc ii} $\lambda$4686)      &0.005$\pm$0.000 \\
He/H                                &0.081$\pm$0.001 \\ \hline
%$Y$                                 &0.245$\pm$0.004  \\ \hline     
\end{tabular}

$^{\rm a}$ICF is the ionization correction factor.
\end{center}

\end{table}
%---------------------------------------------------------------------------------------------

%____________________________________________________________________
\section{Chemical abundances \label{chem}}
%--------------------------------------------------------------------

The observed fluxes of the emission lines have been corrected for 
underlying stellar absorption (for hydrogen lines) and interstellar 
extinction using the observed Balmer decrement
as described by Izotov, Thuan \& Lipovetsky (\cite{ITL94,ITL97}).
The corrected emission line fluxes $I$($\lambda$) relative to the 
H$\beta$ emission line flux, their equivalent widths $EW$, the extinction 
coefficient $C$(H$\beta$), the observed flux of the H$\beta$ emission line, 
and the equivalent width of the hydrogen absorption lines
for the brightest H {\sc ii} region are shown in Table \ref{t:Intens}.

To derive element abundances we adopted a 
spherically symmetric ionization-bounded H {\sc ii} 
region model (Stasi\'nska \cite{S90}) including a high-ionization zone 
with temperature $T_{\rm e}$(O {\sc iii}), and a low-ionization zone with 
temperature $T_{\rm e}$(O {\sc ii}).  
The electron temperature $T_{\rm e}$(O {\sc iii}) is derived from the 
[O {\sc iii}]$\lambda$4363/($\lambda$4959+$\lambda$5007) ratio 
using a five-level atom model. That temperature is used for the 
derivation of the He$^{+}$, O$^{+2}$, Ne$^{+2}$ and Ar$^{+3}$ ionic 
abundances. The electron temperature in the inner He$^{+2}$ zone is expected
to be higher than $T_{\rm e}$(O~{\sc iii}). However, the dependence on the
temperature of the He {\sc ii} $\lambda$4686 emissivity is weak
(Aller \cite{Aller84}). Therefore, $T_{\rm e}$(O {\sc iii}) is adopted 
for the He$^{+2}$ ionic abundance determination.
We derive $T_{\rm e}$(O {\sc ii}) from the relation between
$T_{\rm e}$(O {\sc ii}) and $T_{\rm e}$(O {\sc iii}) (Izotov et al. \cite{ITL94}), 
based on a fit to the photoionization models of Stasi\'nska (\cite{S90}). The 
temperature $T_{\rm e}$(O {\sc ii}) is 
used to derive the O$^+$ and S$^+$ ion abundances. For Ar$^{+2}$
and S$^{+2}$ we have adopted an electron temperature intermediate between
$T_{\rm e}$(O {\sc iii}) and $T_{\rm e}$(O {\sc ii}) following the 
prescriptions of Garnett (\cite{G92}). The electron number density 
$N_{\rm e}$(S {\sc ii}) is derived
from the [S {\sc ii}] $\lambda$6717/$\lambda$6731 flux ratio.
The oxygen abundance is O = O$^{+}$ + O$^{+2}$ + O$^{+3}$, where O$^{+3}$ is
derived from the equation
O$^{+3}$/(O$^{+}$+O$^{+2}$) = He$^{+2}$/He$^{+}$. 
Total abundances of other heavy elements were computed after correction for 
unseen stages of ionization as described in Izotov et al. (\cite{ITL94}) and 
Thuan et al. (\cite{til95}). Five He {\sc i} emission lines $\lambda$3889,
$\lambda$4471, $\lambda$5876, $\lambda$6678, $\lambda$7065 were used
to derive the electron number density $N_{\rm e}$(He {\sc ii}) and the
optical depth $\tau$(He~{\sc i}~$\lambda$3889), and to correct the He {\sc i} 
emission line fluxes for the collisional and fluorescent enhancements
according to Izotov et al. (\cite{ITL94,ITL97}). The singly ionized
helium abundance He$^{+}$/H$^{+}$ is derived as a weighted 
mean of the abundances obtained from the corrected
He {\sc i} $\lambda$4471, $\lambda$5876, $\lambda$6678 line fluxes. 
The total helium abundance is He/H = He$^{+}$/H$^+$ + He$^{+2}$/H$^+$.
The electron temperatures $T_{\rm e}$(O {\sc iii}), $T_{\rm e}$(S {\sc iii}),
$T_{\rm e}$(O {\sc ii}) for the high-, intermediate- and low-ionization 
regions respectively, the electron number 
densities $N_{\rm e}$(S {\sc ii}), ionization correction factors (ICF), 
ionic and total heavy element abundances 
%The derived parameters 
are shown in Table \ref{t:Chem}.

Note that the determination of $N_{\rm e}$ from the 
[S~{\sc ii}]~$\lambda$6717, 6731\AA\ emission lines and of sulfur abundance 
from the [S {\sc iii}] $\lambda$6312\AA\ emission line is 
not very accurate because of the low spectral resolution of the spectrum. 
The spectral resolution is too low to resolve the H$\alpha$ $\lambda$6563\AA\
and [N {\sc ii}] $\lambda$6583\AA\ emission lines. Therefore, the nitrogen
abundance was not derived.

The oxygen abundance 12 + log(O/H) = 7.56 $\pm$ 0.01 
%($Z_\odot$/23)
for the brightest H {\sc ii} region is 
in fair agreement
with 12 + log O/H = 7.52 $\pm$ 0.01 derived by Fricke et al. 
(\cite{Fricke01}) and 7.54 $\pm$ 0.01 derived by Izotov et al. (\cite{ICG01}).
Despite the uncertainties caused by the low spectral resolution, the abundance
ratios Ne/O, S/O and Ar/O are in good agreement with the 
previous abundance determinations in Tol 1214--277 and mean values obtained 
for a sample of the most metal-deficient BCDs (Izotov \& Thuan \cite{IT99}). 
%However, the
%Ne/O abundance ratio in Tol 1214--277 appears by $\sim$ 0.15 dex lower than 
%the mean value for the most metal-deficient BCDs and may suggest that the
%ionization correction factor $ICF$(Ne) = (O$^+$+O$^{2+}$)/O$^{2+}$ used for
%the neon abundance determination in this paper and in Izotov \& Thuan 
%(\cite{IT99}) is not very precise and may overestimate the neon abundance
%in the H {\sc ii} regions with low O$^{2+}$/O$^+$ 
%(see Izotov et al. \cite{ISGT03}). Here O$^+$ and O$^{2+}$ 
%are the abundances of oxygen ions. The use of the ICF(Ne) from Izotov et al.
%(\cite{ISGT03}) almost eliminates the difference between the Ne/O abundance
%ratio in Tol 1214--277 and the mean value of Ne/O abundance ratio for the 
%BCDs from Izotov \& Thuan (\cite{IT99}).

\section{High-ionization emission lines}

Our observations confirm the presence of the [Fe~{\sc v}]~$\lambda$4227\AA\
and the strong He {\sc ii} $\lambda$4686\AA\ emission lines (Fig.
\ref{f:bright}, Table \ref{t:Intens}). Furthermore, we have detected 
for the first time in a star-forming galaxy the [Ne {\sc v}] 
$\lambda$3426\AA\ and marginally the [Ne {\sc v}] $\lambda$3346\AA\ emission 
lines. The [Ne~{\sc v}]~$\lambda$3426\AA\ emission line is seen in
all four spectra of Tol 1214--277 obtained during two nights.
Its width in the averaged spectrum is smaller than that of other emission 
lines, although the line is broader in the spectrum obtained during the
second night. This difference is apparently due to the
noisy spectrum in the blue part and the weakness of the 
[Ne {\sc v}] $\lambda$3426\AA\ line.
% However, at the 68\% confidence level 
%the difference in the line widths is not significant.}

To produce strong [Ne~{\sc v}] $\lambda$3426\AA\ 
the ionizing radiation must be intense at $\lambda$ $\la$ 128\AA, because
the ionization potential of the Ne$^{+4}$ ion is 7.138 Rydberg.
Using the flux ratio 
$I$($\lambda$3346+$\lambda$3426)/$I$(H$\beta$) from Table \ref{t:Intens} and 
expressions from Aller (\cite{Aller84}) we can estimate the fraction of
the Ne$^{+4}$ ions in the H {\sc ii} region.
This ion exists in the inner He$^{+2}$ zone where the electron temperature 
is expected to be higher than that derived from the O {\sc iii} forbidden
lines. To derive the electron temperature and other physical parameters
 in the inner zone of Tol 1214--277, 
we use the CLOUDY code 
%(Ferland \cite{F96}; 
(version C94.00; Ferland et al.
\cite{F98}). Adopting the effective temperature of the ionizing radiation
$T_{\rm eff}$ = 50000 K and CoStar stellar atmosphere models (Schaerer 
\& de Koter \cite{S97}), we obtain 
$T_{\rm e}$(Ne {\sc v}) = 36000 K or 
17000 K higher than $T_{\rm e}$(O {\sc iii}) = 19400 K
(Table \ref{t:Chem}). Then  Ne$^{+4}$/H$^+$ = 8.3$\times$10$^{-8}$ or 
$\sim$ 2\% of the Ne$^{+2}$ abundance.

We used Kurucz (Kurucz \cite{K79}) and CoStar stellar atmosphere models and
calculated several spherically symmetric ionization-bounded 
H {\sc ii} region models which 
reproduce resonably well the observed emission line fluxes of the O$^+$, 
O$^{+2}$, Ne$^{+2}$, S$^{+2}$, Ar$^{+2}$ and Ar$^{+3}$ ions. However,
the observed [Ne {\sc v}] $\lambda$3426/[Ne {\sc iii}] $\lambda$3868 
flux ratio in Tol 1214--277 
is $\sim$ 10$^4$ times higher than that predicted by the 
H {\sc ii} region models even in the case of 
the CoStar atmosphere models of the hottest main-sequence stars
(model F1 with $T_{\rm eff}$ $\approx$ 54000 K;
Schaerer \& de Koter \cite{S97}). 
The difference between the observations and
model predictions is even larger when Kurucz stellar 
atmosphere models are used.

We now estimate the X-ray luminosity required to reproduce the
large observed intensity of Ne {\sc v} emission. The H$\beta$ luminosity  
$L$(H$\beta$) $\approx$ 2.5$\times$10$^{40}$ erg s$^{-1}$ in Tol 1214--277
corresponds to a number of ionizing photons log Q$_{\rm H}$ 
$\approx$ 52.7 (Q$_{\rm H}$ is in s$^{-1}$). 
The derived value of log Q$_{\rm H}$ is a lower limit. It might be higher 
if, e.g., the
H {\sc ii} region is density-bounded or the gas does not fully cover the
ionizing source. Additionally, dust might be present. However,
the density-bounded H {\sc ii} region models are likely excluded because
the observed flux of the [O {\sc ii}] $\lambda$3727 emission line is not 
reproduced by those models. We cannot exclude the presence of dust. 
However, its amount in Tol 1214--277 is likely small, as evidenced by the
strong Ly$\alpha$ emission line (Thuan \& Izotov \cite{ti97}). 
Scaling the CoStar model F1 (Schaerer \& de Koter \cite{S97})
to the derived Q$_{\rm H}$ value, we obtain a predicted X-ray 
luminosity $L_{\rm x}^{\rm mod}$($\lambda$ $\la$ 128\AA) 
$\approx$ 10$^{35}$ -- 10$^{36}$ erg s$^{-1}$. However,  
since the observed [Ne {\sc v}] $\lambda$3426/[Ne {\sc iii}] $\lambda$3868
flux ratio is $\sim$ 10$^{4}$ times larger than the model prediction,
the soft X-ray luminosity of Tol 1214--277 should be
as high as $L_{\rm x}^{\rm obs}$ = 10$^{39}$ -- 10$^{40}$ erg s$^{-1}$ 
to reproduce the observations. 

Tol 1214--277 has not been observed in the X-ray range. However,
the estimated $L_{\rm x}^{\rm obs}$ is consistent with the 0.5 -- 10 KeV 
(or 1 -- 25\AA) X-ray luminosity 
derived from {\it Chandra} observations of another BCD with high-ionization 
lines, SBS 0335--052 (Thuan et al. \cite{T04}).
The large difference between 
$L_{\rm x}^{\rm obs}$ and $L_{\rm x}^{\rm mod}$ 
in Tol 1214--277 implies that the dominant source of soft X-ray emission 
cannot be normal massive main-sequence stars. 
Note that this source is likely to be compact and located in the inner
part of the H {\sc ii} region to produce X-ray emission with a high
ionization parameter.
Fast shocks with the velocities of $\sim$ 400 -- 500
km s$^{-1}$, contributing up to 10\% of the H$\beta$ luminosity, can be 
responsible for the observed high fluxes of both 
the He {\sc ii} $\lambda$4686\AA\
and [Ne {\sc v}] $\lambda$3346, 3326\AA\ emission lines (Dopita \& 
Sutherland \cite{DS96}). Another possibility is 
the ionizing radiation produced by 
the accretion of gas in high-mass X-ray binaries.
This mechanism appears to be at work in SBS 0335--052 
(Thuan et al. \cite{T04}). If additional mechanisms of ionization and 
heating are important in the H {\sc ii} region of Tol 1214--277 then the 
[O {\sc iii}] $\lambda$4363 emission line may be enhanced resulting in
the higher temperature $T_{\rm e}$(O {\sc iii}). In this case the heavy 
element abundances shown in Table \ref{t:Chem} may be underestimated.

%_____________________________________________________________________________
 \section{Summary \label{conc}}
%-----------------------------------------------------------------------------

New spectroscopic observations of the blue compact dwarf (BCD) galaxy 
Tol 1214--277 are
presented which reveal for the first time in a star-forming galaxy 
the strong ($\sim$ 2\% of H$\beta$)
[Ne {\sc v}] $\lambda$3426\AA\ and probably 
the [Ne {\sc v}] $\lambda$3346\AA\  
high-ionization lines. This finding implies the presence of intense
ionizing X-ray emission in Tol 1214--277 from one or several compact sources 
which are not the usual massive stars. Other mechanisms, such as
fast shocks or gas accretion in 
high-mass X-ray binaries, can probably account for the 
large luminosity in the [Ne {\sc v}] $\lambda$3426\AA\ emission line. 
Because its metallicity is low and star-forming activity is high, 
Tol 1214--277 may be a good approximation to 
primordial galaxies, and its study can shed light on 
the physical conditions of high-redshift
galaxies. In particular, we expect relatively strong high-ionization 
emission lines to be present in the spectra of primordial galaxies. 

\begin{acknowledgements}
Y.I.I. acknowledges the G\"ottingen Academy of Sciences
for a Gauss professorship. He and N.G.G. have been supported by DFG grant 
436 UKR 17/22/03 and by  
Swiss SCOPE 7UKPJ62178 grant. They are grateful for the hospitality 
of the G\"ottingen Observatory. 
%K.G.N. acknowledges the support of the Deutsche 
%Forschungsgemeinschaft (DFG) grant FR325/50-1.
%Y.I.I., K.G.N, N.G.G., 
P.P. and K.J.F. acknowledge support by the Volkswagen 
Foundation under grant No. I/72919. Y.I.I. and T.X.T. have been partially 
supported by NSF grant AST 02-05785. Y.I.I., N.G.G. and T.X.T. acknowledge the
support by the grant UP1-2551--KV-03 of the U.S. Civilian Research and
Development Foundation (CRDF) for the independent states of the former Soviet
Union.
%Research by P.P. and K.J.F. has been supported by the
%Deutsches Zentrum f\"{u}r Luft-- und Raumfahrt e.V. (DLR) under
%grant 50\ OR\  9907\ 7. 
\end{acknowledgements}

%------------------------------------------------------------
{}
\end{document}